\documentclass[aps,prl,preprint,floatfix,superscriptaddress]{revtex4}
\usepackage{latexsym} \usepackage{amsmath} \usepackage{dcolumn}
\usepackage{bm} \usepackage{graphicx} \usepackage{epsfig}
 \newcommand{\sgn}{{\rm sgn}\,}

\begin{document}

%\preprint{}

\title{ Many-pole model of inelastic losses in x-ray absorption spectra}



%\thanks{}

\author{J. J. Kas} \affiliation{Dept.\ of Physics, Univ.\ of
Washington Seattle, WA 98195}

\author{A. P. Sorini} \affiliation{Dept.\ of Physics, Univ.\ of
Washington Seattle, WA 98195}

\author{M. P. Prange} \affiliation{Dept.\ of Physics, Univ.\ of
Washington Seattle, WA 98195}

\author{L. W. Cambell} \affiliation{Pacific Northwest National Lab.\
Richland, WA 99352}

\author{J. A. Soininen} \affiliation{Div. of X-ray Physics, Dept. of
Physical Sciences, Univ.\ of Helsinki, Helsinki FI-00014, FI}

\author{J. J. Rehr} \affiliation{Dept.\ of Physics, Univ.\ of
Washington Seattle, WA 98195}


\date{\today}

\begin{abstract}
%Structure refinement via x-ray absorption fine structure (EXAFS) analysis
%relies on effiecient calculation of the quasiparticle self-energy over
%a broad energy range. To date,
%practical calculations over this energy range have been
%phenomenological in nature. Further many body effects are lumped into
%nuisance parameters such as the many body amplitude reduction
%factors $S_0^2$.
Inelastic losses are crucial to a quantitative analysis of x-ray absorption
spectra. However, current treatments are semi-phenomenological in nature.
Here a first-principles, many-pole generalization of the plasmon-pole model
is developed for improved calculations of inelastic losses.
%in x-ray absorption spectra (XAS).
The method is based on the GW approximation for the self-energy and
real space multiple scattering calculations of the dielectric
function for a given system. %$\epsilon(\bm{q},\omega)$
%in the long-wavelength limit for a given system.
%($\bm{q}=0$)
%and a subsequent extrapolation to finite momentum transfer.
The model retains the efficiency of the plasmon-pole model and
is applicable both to periodic and aperiodic materials over a wide
energy range.
%including the near edge spectra.
%in terms of a many-pole approximation.
%to the loss function
%$-\rm{Im}[\epsilon(\bm{q}=0,\omega)^{-1}]$. 
%The self-energy is then obtained using an extension of the single-pole
%GW approximation. 
The same many-pole model is applied to extended GW
calculations of the quasiparticle spectral function. This yields
estimates of multi-electron excitation effects, e.g., 
the many-body amplitude factor $S_0^2$ due to intrinsic losses.
%$S_0^2$ in the fine structure.
Illustrative calculations are compared
with other GW calculations of the self-energy,
the inelastic mean free path, and experimental x-ray absorption spectra.
%and shown to be an improvement over the single-pole self-energy.
\end{abstract}

\pacs{}
\keywords{XANES, EXAFS, PP, MPSE, IMFP}

\maketitle

\section{Introduction}
\label{IntroSect}

The theory of the extended x-ray absorption fine structure (EXAFS) is now
well developed and can be calculated quantitatively in many
systems. \cite{rehr00}
However, x-ray absorption near edge structure (XANES) calculations
have remained only semi-quantitative at best.\cite{rehr05} 
% For x-ray absorption spectra (XAS) and electron energy loss spectra (EELS),
%correspond to the phenomenological definitions of the XANES and EXAFS regions.
One of the reasons for this disparity is a lack of accurate treatments of
inelastic losses in the near edge region.
For example, traditional calculations of EXAFS typically rely on
simplified or semi-phenomenological models of inelastic losses in terms
of a complex, energy-dependent exchange-correlation potential, 
i.e., the quasi-particle self-energy $\Sigma(E)$, where
$E$ is the quasi-particle energy.  In addition
a many body amplitude factor $S_0^2$ must be applied to
the EXAFS signal to account for intrinsic losses, though this is frequently
ignored or considered to be a free parameter.\cite{rehr05} Two
commonly used models for the self-energy in x-ray absorption spectra
(XAS) are 
%based on the electron gas, models
i) the Hedin-Lundqvist plasmon-pole model, and ii)
the Dirac-Hara exchange approximation plus a constant
complex potential.  \cite{Leon_91,rehr00,rehr06,GNXAS_95,FDM_2001}
%to incorporate the effects of
%the self-energy and inelastic losses on the spectra.
Since the self-energy is smoothly varying at high
energy and relatively small compared to the kinetic energy,
these approximations are often adequate for EXAFS. 
%In addition
%a many body amplitude factor $S_0^2$ must be applied to
%the EXAFS signal to account for intrinsic losses. \cite{rehr05}
% Indeed, in traditional analyses
%of EXAFS, the addition of a constant imaginary potential and
%shift of the Fermi energy may be adequate to correct the errors in
%approximate models.
However, variations in the self-energy tend to be large 
in the XANES region, i.e., within the first 50 eV above the Fermi energy,
and neither of the above models describes this variation correctly.
The energy scale mentioned above is set by the dominant excitations in the
system, and is comparable to the mean plasma frequency
$\omega_p$, which is typically about 10-30 eV. 
Thus the EXAFS (characterized by weak scattering due to large loss)
and XANES (characterized by large scattering and low loss) regions
correspond to low and high energy relative to $\omega_p$.
%since the self-energy lacks structure in the region of interest,
As a result, the variation in $\Sigma(E)$ with energy leads to
significant errors both in amplitude and peak positions in the XANES.

In an effort to improve on these simplified models
%current treatments of these many-body effects
%the self-energy and inelastic losses,
we present here a many-pole GW approximation for the self-energy,
\cite{hedin69,aryasetiawan_98,inkson_84} 
based on real space multiple-scattering calculations of the
inverse dielectric function for a given system. % as described below.
Our goal is to develop an approach which
can be applied routinely both for EXAFS and XANES.
Analogous many-pole models have been used previously in 
calculations of the self-energy,\cite{horsch87} and of the inelastic mean
free path (IMFP),\cite{penn87,penn75} with experimental
optical data as input.  A few first principles approaches
that make use of pole approximations have also
been developed.\cite{engel_91_1,engel_91_2,soininen03}
For example Ref.\ \onlinecite{soininen03} makes use of a
band-Lanczos algorithm to calculate a many-pole approximation to the
inverse dielectric matrix. For reviews
%of first principles and phenomenological
of other approaches to GW calculations
see Refs.\ \onlinecite{aryasetiawan_98} and \onlinecite{albur_2000}.  

Our many-pole model yields semi-quantitative self-energies over a wide range
of photoelectron energies from the near-edge to about
$10^3$ eV, which is adequate to cover both the XANES and EXAFS
regions.\cite{sorini06}
The approach has a number of advantages for practical calculations. 
First the method is computationally efficient in that only a few cpu
hours are required to calculate the dielectric function,
self-energy, and spectral function for a given system.
This is significant since XANES calculations typically take
several cpu-hours, while full
GW self-energy calculations over the complete energy range of XAS
experiments are currently impractical.  Finally,
the approach is applicable to a wide class of materials including metals,
insulators, and molecular systems.

The strategy of our treatment of inelastic losses
%and related many-body effects
is as follows.  We begin with a
first principles calculation of the energy loss spectrum
$L(\omega) = -{\rm Im} [\epsilon(\bm{q}=0,\omega)^{-1}]$
in the long wavelength limit $\bm{q}=0$.
\cite{prangeetal05,prangetables05}  
%The model can be regarded
%as a direct extension of the single plasmon-pole model of Hedin and
Next this loss function is incorporated into a many-pole model for the
self-energy which is an extension of the single plasmon-pole model of Hedin and
Lundqvist.\cite{BIL67_I,BIL67_II,hedin69} This self-energy yields system
dependent extrinsic losses due to the lifetime of the quasi-particle
over a broad energy range.  Next, to account for intrinsic losses, i.e., 
losses due to excitations of the system in response to the sudden
creation of the core hole, we apply our many-pole model to
a calculation of the quasiparticle spectral function using an
extension of the GW approximation based on the quasi-boson
model.\cite{campbell02,Bard_85,Hedin_89}
%With many of the same ingredients and 
%an extension of the GW approximation \cite{campbelletal}
%also calculate the quasiparticle spectral function
%$A(k,\omega)$. 
This yields corrections to the quasi-particle
approximation for XAS in terms of a
%In order to move away from the single particle picture,
convolution of the quasi-particle absorption spectrum with the
spectral function. % $A(k,\omega)$.
%These corrections correspond to the {\it intrinsic
%losses}, i.e., multi-electron excitations of the system created in response
%to the core-hole.
Moreover, the approach naturally includes interference terms between extrinsic
and intrinsic losses and describes the crossover from the adiabatic-
to sudden-approximation limits.

The remainder of this paper is organized as follows. 
%In Sec.\ \ref{SPSESect} 
We begin with a brief description of the single plasmon-pole GW model for
the self-energy, together with our extension to many poles. 
%Sec.\ \ref{DFSect} 
We then describe our approach for calculating the dielectric
function at zero momentum transfer, as well as the extrapolation to
finite momentum transfer. 
%Sec.\ \ref{SEResSect} 
Next we compare
our results for the self-energy and the IMFP with other calculations, as
well as with experimental results for the IMFP. 
%Sec.\ \ref{SFSect} 
Subsequently we present our calculation of the quasiparticle spectral
function and it's relation to the self-energy.  
%Sec.\ \ref{ResultSect} 
We then compare our calculations of XAFS with experimental as well as
theoretical results. Finally, we summarize
our results and discuss possible improvements.
%% The remainder of this paper is organized as follows. In Sec.\
%% \ref{SPSESect} we briefly review the single plasmon-pole GW model for
%% the self-energy, and introduce our extension to a many pole model. Sec.\
%% \ref{DFSect} describes our approach for calculating the dielectric
%% function at zero momentum transfer, as well as the extrapolation to
%% finite momentum transfer. Sec.\ \ref{SEResSect} gives a comparison of
%% our results for the self-energy and the IMFP to other calculations as
%% well as experimental results for the IMFP. Sec.\ \ref{SFSect} presents
%% our calculation of the quasiparticle spectral function and it's
%% relation to the self-energy.  Sec.\ \ref{ResultSect} contains a
%% comparison of our calculations of XAFS with experimental and
%% theoretical results. Finally, Sec.\ \ref{ConcSect} contains a summary
%% of our results and discusses possible improvements.

%\subsection{}
%\subsubsection{}
\section{Many-pole self-energy}
\label{MPSESect}
The many-pole model for the self-energy developed here is an extension
of the plasmon-pole (PP) model of Hedin and
Lundqvist,\cite{BIL67_I,BIL67_II,hedin69} and contains many of the
same ingredients. Thus we begin with a brief description of the PP
model, and subsequently describe the extension to a more
general loss function used in this work. A more detailed
description of the plasmon pole model is given in Appendix \ref{PPApp}.
Throughout this paper all
quantities are given in Hartree atomic units ($e = \hbar = m_{e} = 1$)
unless otherwise noted. 
We begin with the GW approximation for the self-energy\cite{hedin69}
of a homogeneous electron gas
in the momentum representation, 
\begin{equation}
  \label{SPSE_Eq}
  \begin{split}
    \Sigma(\bm{k},E) = &\, i \int \frac{d^3q}{(2
    \pi)^{3}}\,\frac{d\omega}{2 \pi}
    \frac{V(q)}{\epsilon(\bm{q},\omega)}\\ \times
    &\frac{1}{E-\omega-E_{\bm{k}-\bm{q}}
    +i(|\bm{k}-\bm{q}|-k_{F}) \delta},
  \end{split}
\end{equation}
where $k_F$ is the Fermi momentum. In frequency space the imaginary
part of the inverse dielectric function (i.e., the loss function of
the electron gas) is modeled as a single pole at $\omega(q) =
[\omega_p^2 + a q^2 + b q^4]^{1/2}$, where the coefficients of the
dispersion $a=k_{F}^{2}/3$ and $b=1/4$ are chosen following the prescriptions
of Hedin and Lundqvist.\cite{BIL67_II,hedin69} This gives a
single-pole model of the inverse dielectric function where
\begin{equation}
  \label{SPEpsEq}
 -{\rm Im} \left[\epsilon(\bm{q},\omega)^{-1}\right] = \pi \omega_p^2\,
  \delta[\omega^2-\omega(q)^{2}],
\end{equation}
and
\begin{equation}
  {\rm Re} \left[\epsilon(\bm{q},\omega)^{-1}\right] =  1 +
  \frac{\omega_p^2}{\omega^2-\omega(q)^2}.
\end{equation}
Inserting these results into Eq.\ (\ref{SPSE_Eq}) then yields two
terms: the first term can be integrated analytically and gives a
static Hartree-Fock exchange potential $\Sigma_{HF}$ which, in the
local density approximation (LDA) is termed Dirac-Hara exchange,
\begin{equation}
  \Sigma_{HF}(k) = -\frac{k_F}{\pi}\left[1 + \frac{k_F^2 - k^2}{2 k
      k_F}\ln\left|\frac{k_F+k}{k_F-k}\right|\right].
\end{equation}
The second term, denoted by $\Sigma_d({\bm k},E;\omega_p)$, is the
dynamically screened exchange-correlation contribution. This
contribution arises from the creation of virtual
bosons. The integrals over frequency and solid angle can be performed
analytically\cite{BIL67_II}, leaving an expression for 
$\Sigma_d({\bm k},E;\omega_p)$ in terms of a single integral over
momentum transfer $\bm{q}$.
This formulation has been used extensively to calculate the mean
self-energy $\Sigma(E)$ within the LDA 
over a broad range of energies for EXAFS spectra.\cite{rehr00}
The PP approximation works well at high energies
and more generally for systems with sharp plasmon-peaks in the 
inverse dielectric function (e.g., Al),
which can be described by nearly
free electron gas models.
On the other hand the model often loses
accuracy at low energies for transition metals, insulators and
molecules with more complex loss spectra, and in practice
often gives unphysical structure to the self-energy near $\omega_p$.
\cite{rehr00}

In order to improve on the plasmon-pole approximation, we now
introduce a more realistic representation for the inverse dielectric
function, using a sum over discrete poles. This representation preserves
the analytical character of $\epsilon(\omega)^{-1}$, and
corresponds to a distribution of bosonic excitations describing
the dielectric response of a material, including both interband
and intraband excitations. The inclusion of this excitation spectrum
in the self-energy naturally broadens the single PP model in a
way which is characteristic of a given system. Moreover, the representation
can be systematically improved.
\begin{figure}
  \label{excfig}
  \includegraphics[height=\columnwidth, angle=-90]{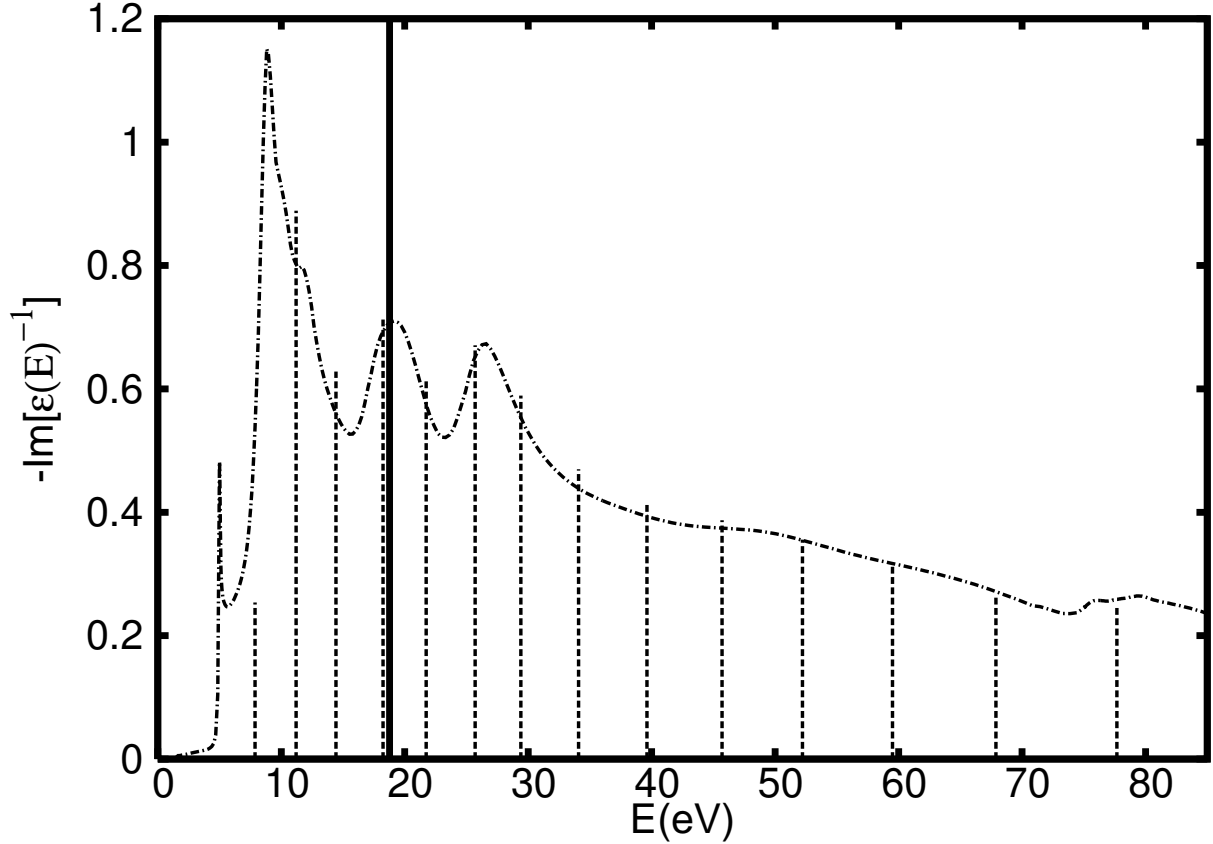}
  \caption{Energy loss function $L(\omega)=-{\rm Im} [\epsilon(\omega)^{-1}$]
in the long wavelength limit for Cu modeled either as
a single pole (solid vertical line), or as a
sum of weighted poles (vertical dashes), compared to the loss function as
calculated by the FEFF8 code (dot dashed).}
\end{figure}

Two steps must be accomplished in order to extend the
PP self-energy to a many-pole self-energy (MPSE): i) The first step
is to obtain a suitable approximation to the energy loss function
$ L(\omega)= - {\rm Im} [\epsilon(\bm{q}=0,\omega)^{-1}]$.
This can be done
either by calculation, as is done here, or from experimental optical
constants. ii) The second step is to extend the ${\bm q}=0$ result to
finite momentum transfer, by representing it as a weighted sum of
poles, each of the form Eq.\ (\ref{SPEpsEq}), which together
conserve the overall strength. 
In addition we approximate the single particle Green's function  $G(E)$
as that for a free particle.  This is the first term in the
multiple-scattering expansion and ignores fine structure;
hence the calculated self-energy represents a uniform average.
With these conditions
the net self-energy is simply the Hartree-Fock exchange contribution plus a
dynamically screened exchange-correlation contribution which is given
by a weighted sum of single pole terms,
%[{\bf NOTE: comment phenomenological models for eps**-1 have been
%used previously, e.g., by Horsch and von der Linden, Fernandez-Varea et al.}]
\begin{equation}
  \Sigma_d(k,E) = \sum_{i}{g_{i}\Sigma_d(k,E;\omega_{i})}
    \label{mpseEq}
\end{equation}
with weights $g_{i}$ and plasma frequencies $\omega_{i}$.
As mentioned above, $\Sigma_d(k,E)$ is given by a single integral over
momentum transfer $|\bm{q}|$, making the calculation quite
efficient. Fig.\ \ref{excfig} illustrates the self energy from
our many pole model for Cu. Note
that only 20 poles were needed to converge this calculation, despite
the relatively broad loss function of Cu.

\subsection{Inverse dielectric function}
\begin{figure} 
  \includegraphics[height=\columnwidth, angle=-90]{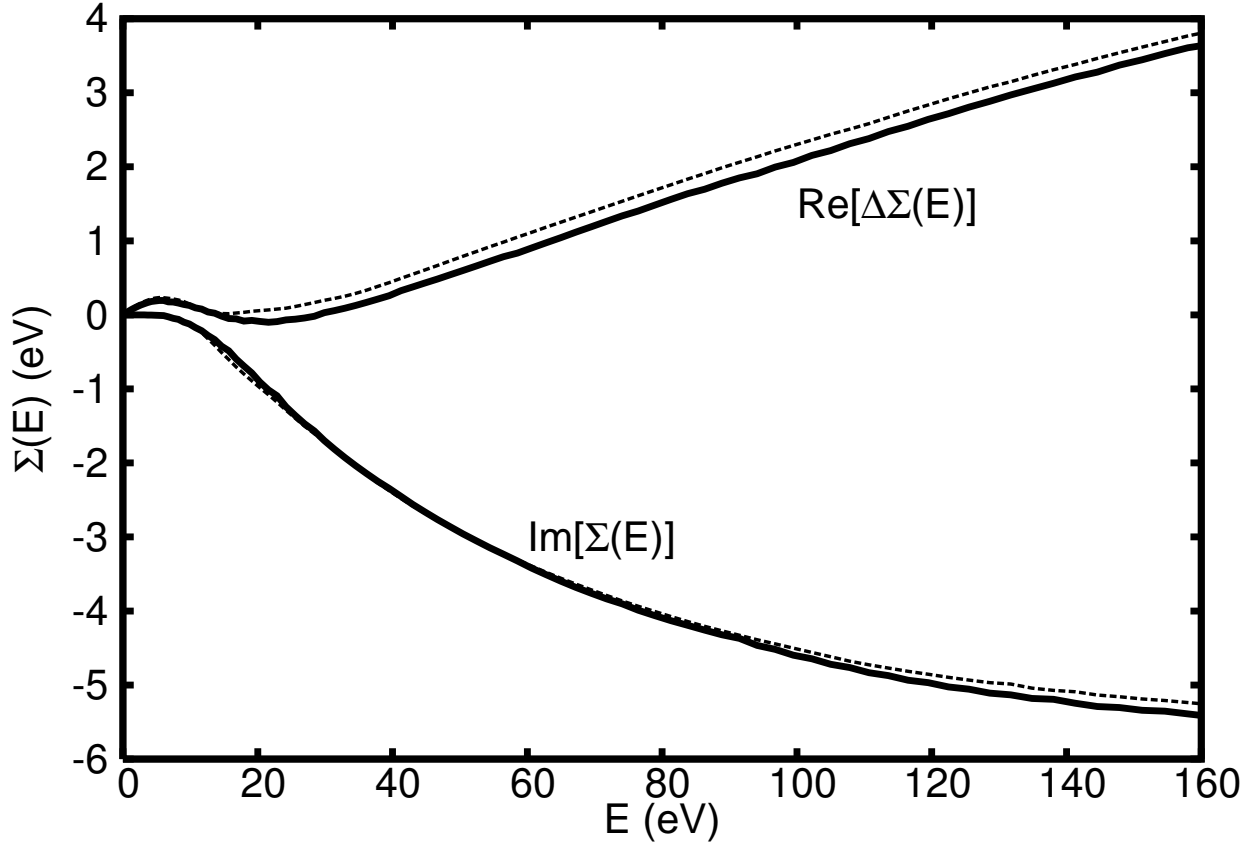}
  \caption{Quasiparticle self-energy for Cu calculated using a loss
   function with fine structure (dashes) or without fine structure
   (solid).}
  \label{FSfig}
\end{figure}

%at zero momentum transfer}
In our approach the inverse dielectric function is calculated  using
the real-space Green's function (RSGF) method as follows:
First, a modification to the RSGF code
FEFF8\cite{feff82ref,feff84ref} is used to calculate the total
absorption cross section $\sigma(\omega)$ for a given material over
a broad spectrum, by summing the contributions from all occupied
initial states.\cite{rehr06,prangetables05} The results presented in
this paper make use of atomic initial states. However current
developments allow for the description of a continuous band of initial
states within the FEFF real space MS framework,\cite{prangeetal05}
which may further improve the results.
%{\sl Josh: do we now want to allow for the initial states being a band ?
%as in the Opcons generalization?}
The imaginary part of the dielectric function $\epsilon_2$ is directly
related to the total absorption cross section per atom $\sigma(\omega)$
as calculated by the FEFF code. $\epsilon_2 = (n/\alpha\omega) \sigma(\omega)$
%\begin{equation}
%%  \rm{Im}[\epsilon(\omega)] = \frac{\it{n}}{\alpha \omega}
%  \epsilon_2(\omega) = \frac{\it{n}}{\alpha \omega}
%    \sigma(\omega),
%\end{equation}
where $\it{n}$ is the atomic number density and $\alpha$ is the fine
structure constant.  The real part $\epsilon_1(\omega)$ is then 
obtained via a Kramers-Kronig transform, and finally
$L(\omega)=-{\rm Im}\,[\epsilon(\omega)^{-1}]$ is formed by inverting
$\epsilon(\omega)$. This could be a computationally demanding
operation. However, because the self-energy involves an 
integral over $\epsilon(\omega)^{-1}$, the fine structure can be
neglected in all
but the lowest energy part of the dielectric function (i.e., the first
20 eV), as shown in Fig.\ \ref{FSfig}.
This approximation considerably reduces the computational effort.
It should be noted that this
prescription for the calculation of the loss function also neglects local
field effects due to the off diagonal components of the dielectric
matrix. Nevertheless, the method has been shown to give reasonable
agreement with experiment for a variety of
materials.\cite{prangeetal05,prangetables05} Moreover, neither the
self-energy nor the absorption spectrum (i.e. EXAFS and XANES) are
highly sensitive to details of the loss function provided the overall
weight is conserved since these quantities are given by integrals over
the loss function.
%In addition, the
%inclusion of local field effects in the calculation of the self-energy
%is not expected to be significant in the final absorption spectrum,
%since the effect of self energy is already second order.
%typically by factor of about 10 corresponding to
%the number of low-lying absorption edges in a material.
%% It must be noted that
%% this prescription for calculating the inverse dielectric function
%% works well at high energies, and gives a fair approximation at low
%% energies for monatomic solids. For molecules, and other multi-atomic
%% systems we must be more carefull in our calculation.
% Change the following.
%% The bulk dielectric function is actually the
%% inverse of the average of $\frac{1}{\epsilon(\bm{r})}$
%% (Cite). Thus for systems that are not homogeneous, excitations from
%% initial states that are localized around each type of atom should be
%% summed and inverted separately.
%% \begin{equation}
%%   \rm{Im}[\epsilon_a] = \sum_{i_a}{\rm{Im}[\epsilon_{i_a}]}
%% \end{equation}
%% \begin{equation}
%%   \rm{Im}[\frac{1}{\epsilon}] = \sum_a{\rm{Im}[\frac{1}{\epsilon_a}]}
%% \end{equation}

\subsection{Extension to finite momentum transfer}
In order to extend the inverse dielectric function to finite momentum
transfer $q$, we represent the imaginary part of the loss function
as a sum of closely-spaced delta functions
\begin{equation}
 L({\bm q},\omega) =
- {\rm Im} \left[\epsilon(\bm{q},\omega)^{-1}\right] = \pi \sum_i{g_i
  \omega_i^2 \delta\left(\omega^2 - \omega_{i}(q)^2\right)}.
\label{epsqw}
\end{equation}
Typically of order $10^1 - 10^2$ poles are sufficient.
Matching the many-pole model $\epsilon(\bm{q},\omega)^{-1}$ evaluated
at zero momentum transfer to the calculation of
$\epsilon(\omega)^{-1}$ then gives the weights ${g_i}$ and pole
locations ${\omega_i}$ respectively.  Our prescription for this match
is as follows: First, the loss function is split into $N$ regions,
each of which is represented by a single pole. For each region of
width $\Delta_{i}$, the pole strength and position are chosen to
preserve first and first-inverse moments, yielding the equations
defining $g_i$ and $\omega_i$
%% \begin{equation}
%%   \int_0^{\omega_{i}} d\omega\, \omega\,
%%   \rm{Im}\left[\epsilon^{-1}(\bm{q},\omega)\right] = \int_0^{\omega_{i}+\Delta_{i}} d\omega
%%   \omega\, \rm{Im}\left[\epsilon^{-1}(\omega)\right]
%% \end{equation}
%% and
%% \begin{equation}
%%   \int_0^{\omega_{i}} \frac{d\omega}{\omega}\,
%%   \rm{Im}\left[\epsilon^{-1}(\bm{q}=0,\omega)\right] = \int_0^{\omega_{i}+\Delta_{i}} \frac{d\omega}
%%   {\omega}\, \rm{Im}\left[\epsilon^{-1}(\omega)\right]
%% \end{equation}
%%are satisfied. Where $\epsilon(\omega)$ is the calculated dielectric
%%function which will be represented from zero to
%%$\omega_{i}+\Delta_{i}$ by the first $i$ poles. Thus the following
%%equations define $g_{i}$ and $\omega_{i}$.
\begin{eqnarray}
  g_{i} \omega_{i}^{2} &=& - \frac{2}{\pi}\int_{\Delta_{i}} d\omega\,
  \omega\, {\rm Im} \left[\epsilon(\omega)^{-1}\right], \\
%\end{equation}
%and
%\begin{equation}
  g_{i} &=& - \frac{2}{\pi}\int_{\Delta_{i}} \frac{d\omega}{\omega}\,
  {\rm Im} \left[\epsilon(\omega)^{-1}\right].
\end{eqnarray}
%which define $g_{i}$ and $\omega_{i}$. 
%Note that this method is quite similar to that of
%Refs. \onlinecite{penn87} and \onlinecite{HV87} where 
For simplicity we also use the same plasmon dispersion
as in the PP model. This approximation has been checked against a
dispersion relation which maintains the width of the pole at high
momentum transfers and gives similar results for materials with a
broad loss function such as Cu, Ag, and Diamond. For materials with a
sharp loss function (i.e. Al, Si, and Na) this approximation may not
be adequate at low energies (below the plasmon energy) where the contribution
from the particle hole continuum can dominate the loss. 

Finally, for very low energies (i.e., the first few eV where our
multiple-scattering calculations are least reliable) the calculated
loss function must be corrected. For metals a Drude term is added and
otherwise a uniform shift of the frequencies $\{\omega_i\}$ is carried
out while scaling the resultant poles, so that the inverse moment
matches either empirical values or accurate calculations of the
static dielectric constant $\epsilon(0)$, while leaving the first
moment unchanged. 
%As a result of these constraints,
%the first moment of $\epsilon_2$ is preserved as well.
%%\begin{equation}
%%  \frac{2}{\pi}\int_{0}^{\infty}\frac{d\omega}{\omega}\,\rm{Im}\left[\epsilon(\bm{q}=0,\omega)^{-1}\right]
%%  = \rm{Re}\left[\epsilon(\omega = 0)^{-1}\right]
%%\end{equation}
%%which gives
%%\begin{equation}
%%  \sum_i{g_i} = 1 - \epsilon(\omega=0)^{-1}
%%\end{equation}
%%In order to do so, we use the following prescription:
%%\begin{enumerate}
%%\item If the integral is too low add a final pole at high energy such that
%%\begin{equation}
%%   -\int_{0}^{\infty} \frac{d\omega}{\omega}\,\rm{Im}\left[\epsilon(\bm{q}=0,\omega)^{-1}\right] = 1 -
%%   \rm{Re}\left[\epsilon(\omega=0)^{-1}\right]
%%\end{equation}
%%
%%\item If the integral is too high, set $\omega_{max}$ such that the
%%  sum of poles is correct.\cite{HV87}
%%\end{enumerate}

For stability we have found it important to preserve the inverse first
frequency moment, since this ensures cancellation of the logarithmic
singularity in the derivative of $\Sigma_{HF}$ at $k=k_F$ and
$E=E_F$.\cite{BIL67_I} This singular behavior otherwise shows up as a
sharp rise in ${\rm Re}\,[\Sigma(\bm{k}(E),E)]$ within the first 10-20 eV
above $E_F$. In metals, where cancellation is perfect, Re
[$\Sigma(\bm{k}(E),E)$] is fairly flat near E=$E_F$. In insulators,
however, this singular behavior is found to enhance the jump in Re
[$\Sigma(\bm{k}(E_F),E_F)$].  Thus our prescription requires a
separate estimate either of the static dielectric function
$\epsilon(0)$ for the case of insulators or the Drude parameters for
metals and semi-metals. These quantities have been used previously
to parameterize the dielectric
matrix,\cite{LLH_82,HyL_86,HyL_88,Cappellini_93,engel_91_1,VLH_88} for
example
Ref. \onlinecite{LLH_82} uses similar parameters to modify a single
pole model of the dielectric function, while
Refs. \onlinecite{HyL_86} and \onlinecite{HyL_88} generalize to a full dielectric
matrix.

Our self-energy model is similar to those of Penn\cite{penn87}
and Horsch et al.\cite{horsch87} for the valence contribution.
One difference is that our formulation neglects the relatively
small particle hole continuum contributions below the plasmon onset.
Another is that our formulation includes a first order
correction to the quasipartle energy as well as a renormalization
constant $Z$ which accounts for the quasiparticle spectral
weight. Appendix \ref{HLQEqApp} gives a short discussion of the
equivalence of the formulas for the self-energy given by Quinn
(which is the starting point for Penn's calculations) and
by Lunqvist. 
A further note must be made regarding the difference between our
many-pole model and the LDA implementation in the
FEFF8.2 code. The plasmon frequency in the current LDA model in FEFF8.2
is dependent on the electron density as a function of spacial coordinates.
In the model discussed here, adding spacial dependence greatly complicates
the theory and is therefore ignored.
%this dependence is ignored.
%the spacial dependence should be delt with,
%since we are using an average loss function.
Thus our approach gives the spacially averaged quasiparticle correction
for the whole system.
We have found that in the XANES region, the
self-energy effects on the spectrum are not sensitive to the
density dependence. Also
our calculations use the interstitial density to determine the Fermi momentum;
the interstitial density was chosen, instead of the average density,
because we want the model to capture the behavior of the self-energy
due to interaction with the valence electrons. For the
core electrons FEFF8 already has an option to use a non-local Dirac-Fock
exchange which can be applied with our many-pole
model.\cite{ankudinov_97}
%\textit{I think we should take the following sentence out alltogether,
%but I gave a possible replacement which follows.}
%[In developing our model other more complicated models were considered,
%including one which reproduces the density dependence of a single-pole
%by scaling the pole positions accordingly, but 
%did not significantly improve the results.]
%Thus a Drude model is used in
%metals to extrapolate to zero frequency, while various methods
%including empirical values
%(i.e. Stott Zaremba, DFT, empirical values, ...)
%may be used to fix $\epsilon(0)$ for insulators.
%%\begin{figure} 
%%  \includegraphics[height=4in, angle=-90]{sing.eps}
%% %%   \includegraphics[width=\columnwidth]{sing}
%% %%   \caption{Re\left[$\Delta\Sigma$\right] for a single pole with different
%% %%   positions. The upper panel shows calculations with pole weight of
%% %%   1.0 which results in full cancellation of the singular behavior near $E=E_F$. The
%% %%   lower panel shows calculations with pole weight 0.5. }
%% %% \end{figure}
\section{Extrinsic Losses}
\label{SEResSect}
\begin{figure}
  \includegraphics[height=\columnwidth, angle=-90]{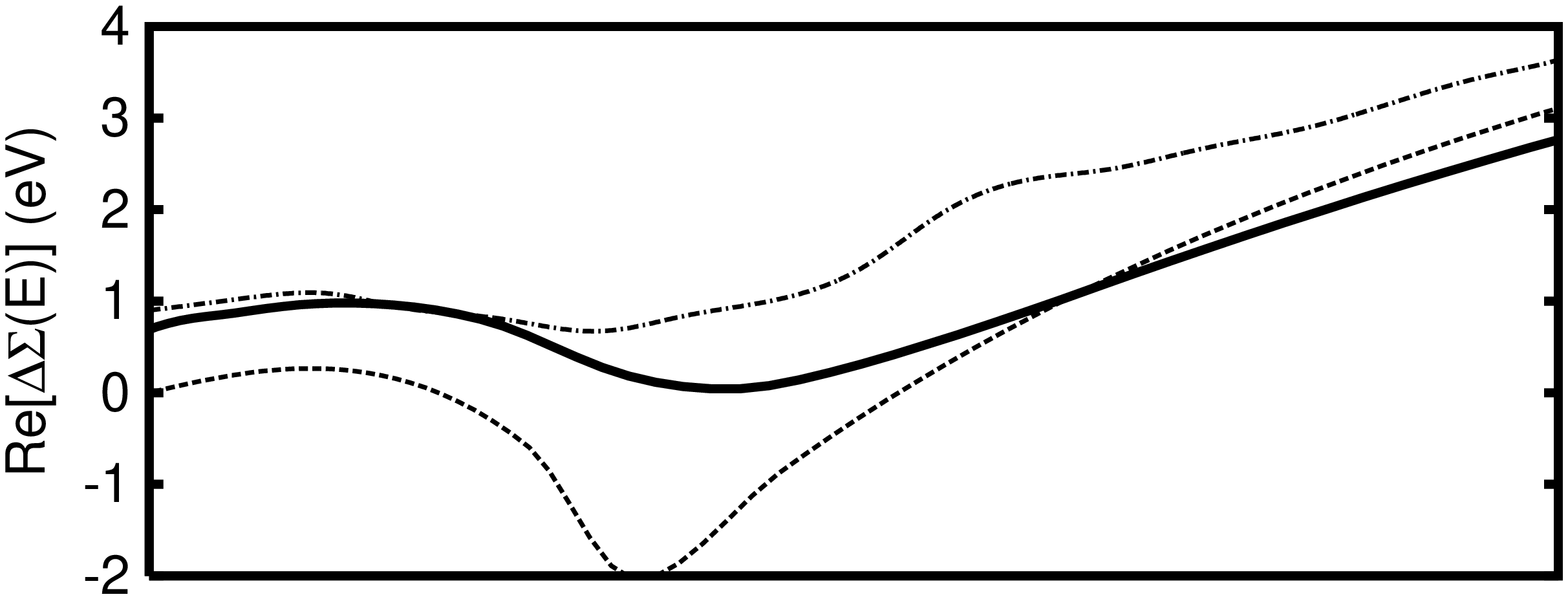}
  \includegraphics[height=\columnwidth, angle=-90]{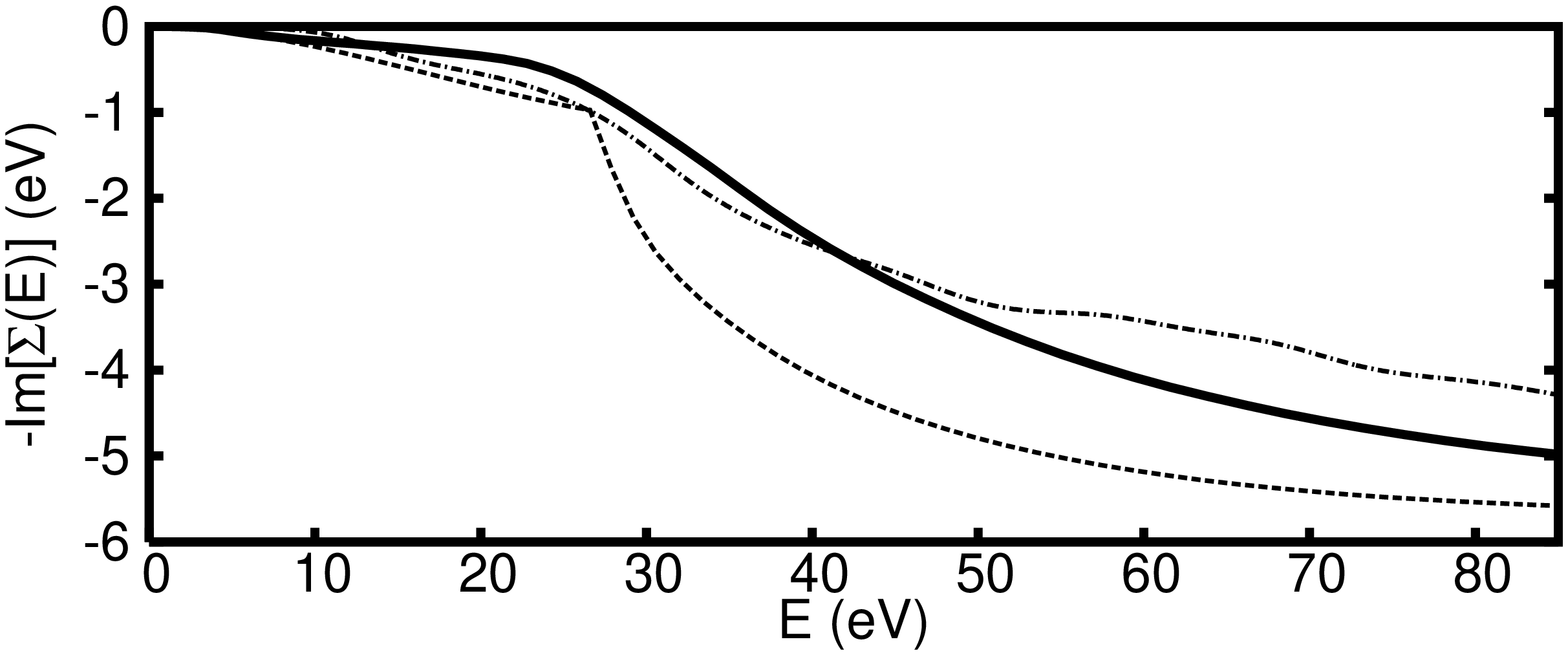}
%  \includegraphics[width=\columnwidth ][width=\columnwidth
%  ]{Diamond.png}
  \caption{a) Real (top) and b) imaginary (bottom) parts of the self-energy
for Diamond calculated using our many-pole model (solid); the
Hedin-Lundqvist PP model (dashes); and the iterative
band-Lanczos calculation of Refs.\ \onlinecite{soininen03} and
\onlinecite{soininen05} (dot dash).}
  \label{csefig}
\end{figure}

\begin{figure}[h]
  \label{IMFP_Fig}
  \includegraphics[height=\columnwidth,angle=-90]{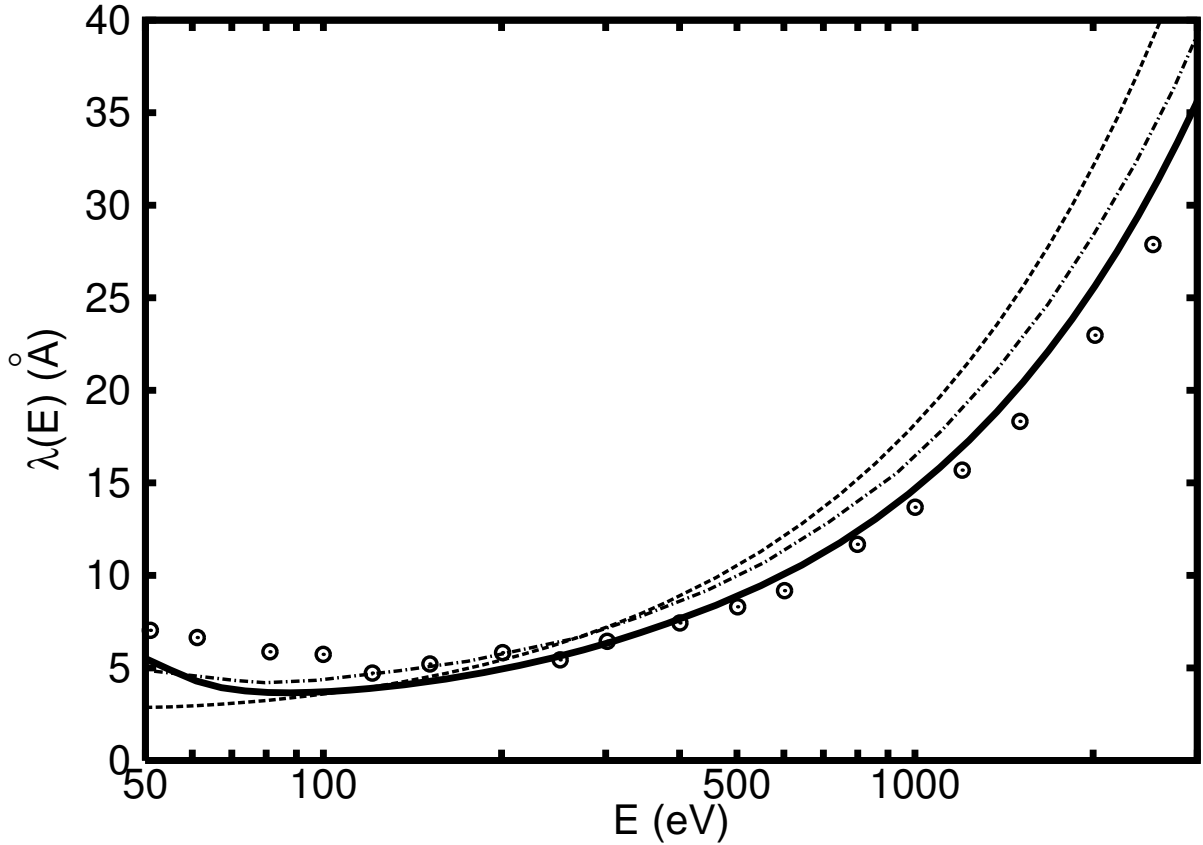}
  \caption{Inelastic mean free path for Mo calculated using our many
pole model (solid); the Hedin-Lundqvist single pole model (dashes); a
many-pole model based on optical data\cite{tanuma05,penn87,penn75} (dot dash);
and experimental data from Tanuma et al.\ \cite{tanuma05} (circles)}
  \label{mofig}
\end{figure}

In this section we present results which characterize the
"extrinsic losses," in XAS, namely the self-energy and the
inelastic electron mean-free-path.
To confirm that our approach gives improved results when compared to
the PP model, we have compared with other calculations of the
self energy, including the single PP model of Hedin and
Lundqvist and a more accurate many-pole
approximation.\cite{soininen03,soininen05} Fig.\ \ref{csefig} shows
our many-pole self-energy for diamond compared with the single-pole
model as well as the band-Lanczos calculation of Refs.\
\onlinecite{soininen03} and \onlinecite{soininen05}. 
%For our many-pole
%calculation, we included a shift of 4.2 eV in order to account for the
%LDA band gap.
In addition, we use our results to calculate the electron inelastic
mean free path (IMFP).
\begin{equation}\label{imfpeq}
  \lambda(E) = \sqrt{\frac{E}{2}}\frac{1}{|{\rm Im} [\Sigma(E)]|}.
\end{equation}
Note that this definition is not the EXAFS IMFP
$\lambda_{\rm EXAFS}$,\cite{rehr00}
since that quantity characterizes the decay of the EXAFS amplitude
and includes both
core hole broadening $\Gamma$ and the self-energy,
$\lambda_{\rm EXAFS} = (2E)^{(1/2)}/ [|{\rm Im} \Sigma(E)]| + \Gamma/2]$.
Fig.\ \ref{mofig} shows our results for the IMFP for Mo, and for
comparison the single-pole model, an optical model which uses the Penn
algorithm,\cite{penn87,penn75} and experiment.\cite{tanuma05}
Other applications of our many-pole model as well as IMFP results for
a number of materials have recently been presented by Sorini et
al.\ \cite{sorini06} As can be seen in Fig.\ \ref{IMFP_Fig}, our
self-energy gives improved results for the IMFP over a broad range of
energies.

% Luke to write
%% The effect of inelastic losses on the x-ray absorption spectrum,
%% although fundamentally resulting from many-body processes, can be
%% represented as a weighted sum (or integral) of energy shifted single
%% particle spectra.  This occurs because the primary determinant of the
%% absorption signal is the photoelectron ejected in the absorption
%% process.  The multiple scattering of the photoelectron off neighboring atoms
%% determines the density of states for the final state, and hence the
%% cross section for absorption of an x-ray\cite{sayers70}.  
\section{Intrinsic Losses}
\label{SFSect}
\label{sect\thesection}

In this section we describe the treatment of intrinsic losses in a
system in terms of an effective quasi-particle ``spectral function." \cite{campbell02} The many-pole GW self-energy developed above is
adequate to describe the extrinsic losses of the photoelectron in the
independent particle (i.e., quasi-particle) approximation for the XAS.
However this approximation neglects intrinsic losses due to the
excitations in the absorbing medium that arise from the sudden
creation of the core-hole. As a consequence of these excitations, the
energy of the absorbing photoelectron is lowered, resulting in a shift
in the absorption signal. 
%For such
%losses, the density of excited states resulting from multiple scattering of the low energy
%photoelectron leads to a single particle absorption spectrum
%of the single particle spectrum
%which is shifted by the energy of the inelastic excitations. 
Moreover, one must also take into account {\it interference} between
the intrinsic and extrinsic losses. Both intrinsic losses and
interference terms can be accounted for in terms of an energy-dependent
spectral function.

Here we implement a many-pole model for the spectral
function derived from a direct extension to the GW approximation 
and based on a quasi-boson model.  \cite{campbell02}
Within the approximations detailed in Ref. \onlinecite{campbell02}, 
the full many-body spectrum is given by a
convolution of the single quasi-particle spectrum with an {\it energy
dependent} spectral function $A_{{\rm eff} }(\omega ,\omega')$, i.e.,
\begin{equation}
  \label{mumb}
  \mu(\omega)=\int \!d\omega'\,A_{\rm eff} (\omega ,\omega')\mu _{\rm
    qp} (\omega -\omega').
\end{equation}
Here $A_{\rm eff} (\omega,\omega')$ characterizes the probability density that a
photon %of frequency $\omega+E_{F}-E_{c}$, where $E_c$ is the core level, 
excites a photoelectron of energy $\omega-\omega'$, as well as
additional excitations (e.g., plasmons, electron-hole pairs, etc.)
with energy $\omega'$.

%% The weight given to each energy in the sum of these energy shifted
%% spectra is known as the \textit{spectral function} $A_{\rm eff}(\omega
%% ,\omega ^{\prime })$.  The variable $\omega'$ is the energy of the
%% inelastic excitations, while $\omega$ is the on-shell kinetic energy
%% of the photoelectron plus the electron self energy.  The many body
%% EXAFS signal $\chi$ is thus given by a convolution of the single
%% particle EXAFS signal $\chi _{\rm sp}$ with $A_{\rm eff}$ (explicitly
%% normalized here) 
Similarly the intrinsic many-body corrections to the EXAFS $\chi$ can
be represented by a convolution of the single quasi-particle signal
$\chi_{\rm qp}$ and the normalized effective
spectral function $A_{\rm eff} (\omega,\omega')$ \cite{campbell02}
\begin{equation}
  \chi(\omega)=\int \!d\omega'\,A_{\rm eff} (\omega ,\omega')\chi _{\rm
    qp} (\omega -\omega').  \label{chimb}
\end{equation}
The convolution in Eq.\ (\ref{chimb}) leads to a {\it path dependent}
amplitude reduction in the EXAFS signal $S_{0,j}^2$.
\begin{figure} 
  \includegraphics[height=\columnwidth,angle=-90]{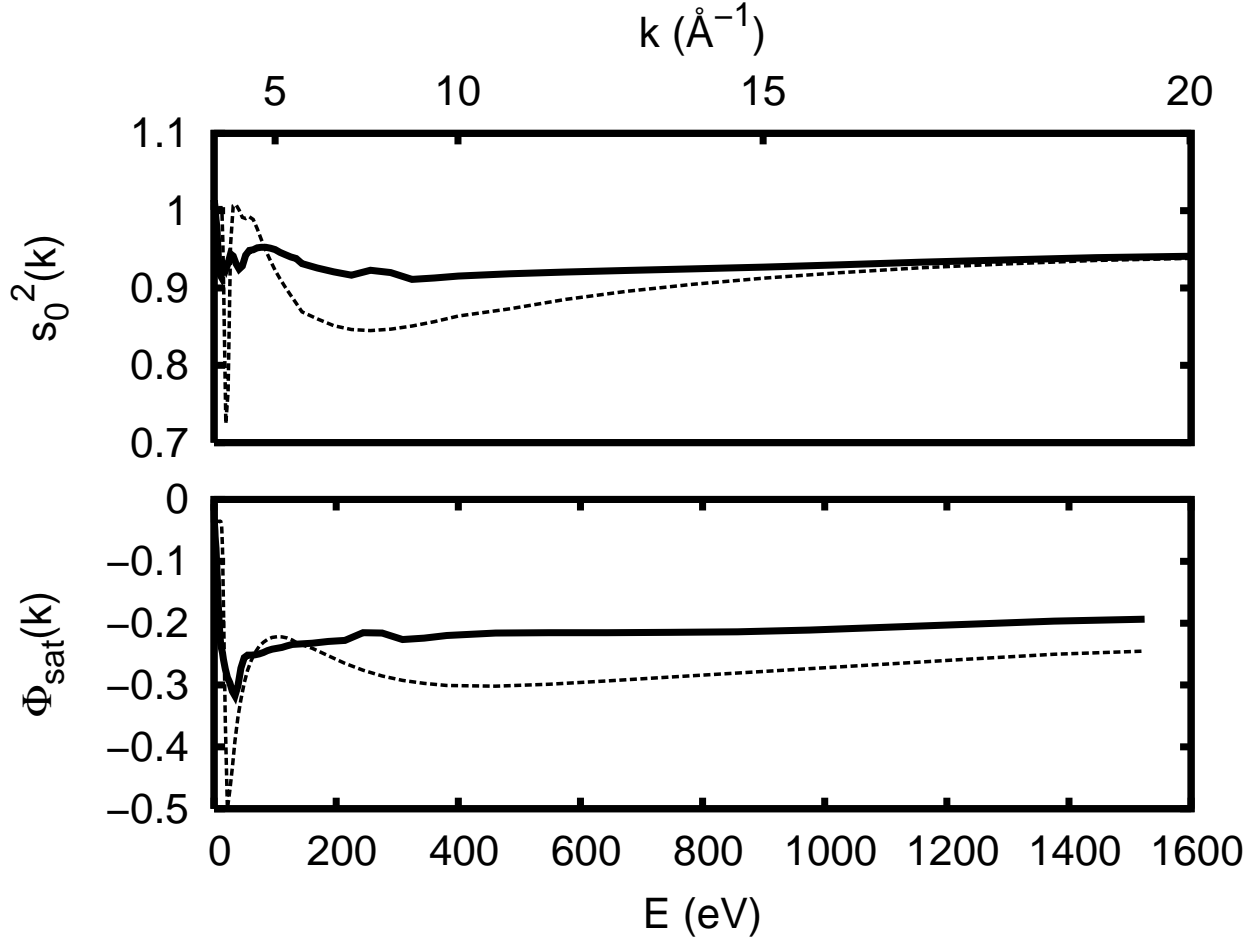}
  \caption{EXAFS $S_{0}^{2}(k)$ (upper) and the net phase $\Phi_{sat}$
of $A_{sat}(k)$ (lower) for our many-pole model (solid) and the
Hedin-Lundqvist single pole model (dashes). Note that the many-body
amplitude and phase factors are approximately constant over a broad
range of energies in the EXAFS ($\approx$ 200 eV and above) with $S_{0}^{2}$
ranging from $\approx 0.91-0.94$ and $\Phi_{sat}$ in the range $-0.21$
to $-0.18$ rad.}
\end{figure}
%= \chi(\omega)/\chi _{\rm qp}(\omega)$. 
Since the EXAFS $\chi$ can be expressed as a sum of rapidly varying
sinusoidal contributions from each photoelectron scattering path\cite{ssl_71} with
smooth amplitudes
%and since for each scattering path $j$ the rapid
%amplitude variation of the signal is sinusoidal 
%\begin{equation}
$\chi_j(\omega) \propto \exp(2iR_j k(\omega))$,
%\end{equation}
the amplitude reduction for each path is given by a phasor-summation
\begin{equation}
  S_{0,j}^2(\omega) \approx \int \!d\omega'\, A_{\rm eff} (\omega
    ,\omega')e^{2iR_{j}\left[k(\omega - \omega')-k(\omega)\right]},
  \label{so2}
\end{equation}
where $R_j$ is one half the total scattering path length of the
photoelectron.  As
$S_{0,j}^2(\omega)$ is only weakly energy dependent, this
amplitude factor can usually be approximated by a constant over a
broad range of energies, consistent with experimental observation.
In contrast the behavior of $S_{0,j}^2(\omega)$ for the
single-pole model exhibits much more variation.

The spectral function can be considered to be made up of a
quasiparticle peak and satellites. Since broadening can be added separately,
the quasiparticle part can be
represented as a delta function of net magnitude $Z_{\rm eff}$ at zero
excitation energy, while the satellites represent contributions from
inelastic excitations in the medium. Within the {\it quasi-boson}
approximation \cite{campbell02} one has
\begin{figure} 
  \includegraphics[height=\columnwidth,angle=-90]{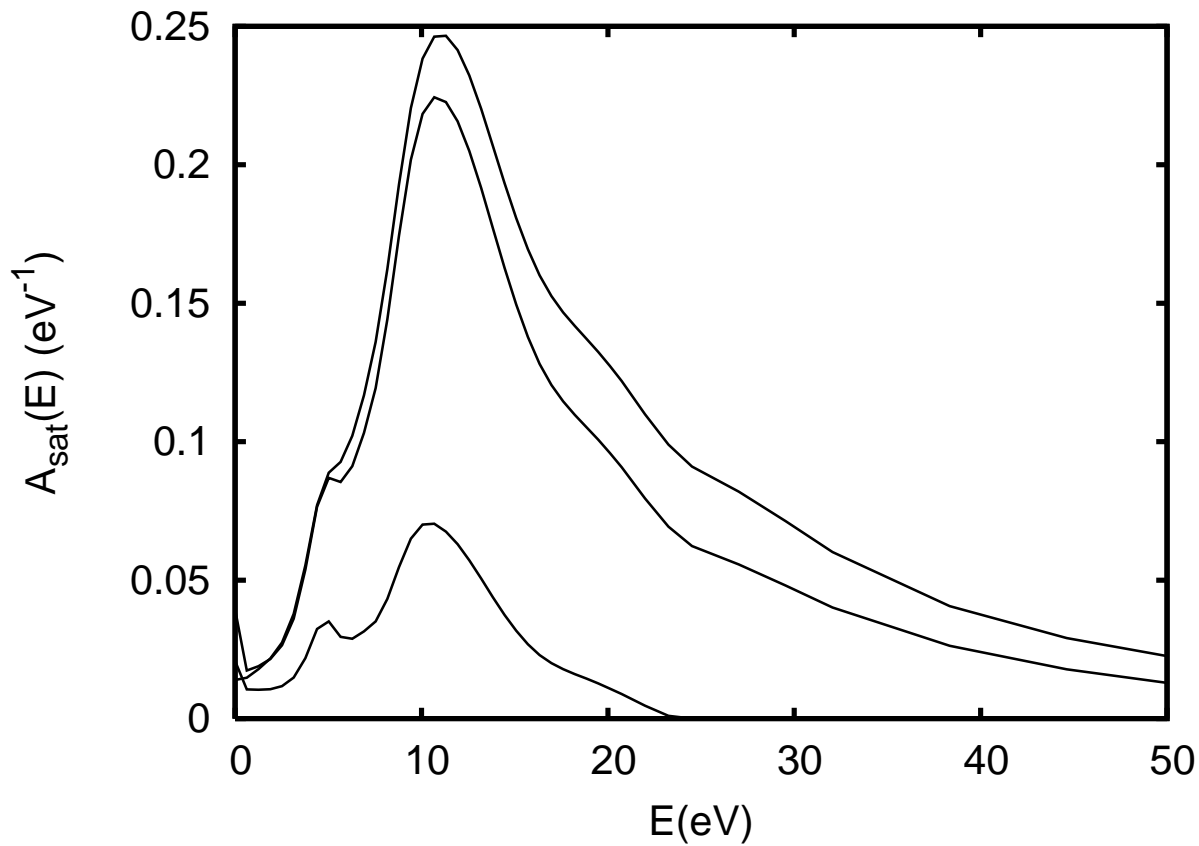}
  \caption{Satellite spectral function for a range of photoelectron
  momenta. From top to bottom, $k=16$, 8, and 4 \AA$^{-1}$.}
\end{figure}
\begin{equation} 
  A_{\rm eff} (\omega ,\omega ^{\prime })=N(\omega)
   \left[\left[1+2a(\omega )\right]\delta (\omega ^{\prime }) +A^{\rm
   sat} (\omega ,\omega ^{\prime })\right],
\end{equation} 
where $N(\omega)$ is a normalization constant which preserves the
overall spectral weight at each $\omega'$. In our approach the
satellite contribution is further broken down into three terms
corresponding to the origin of the inelastic excitation; an
\textit{extrinsic} part $A^{\rm sat}_{\rm ext}$ coming from
excitations created by the photoelectron, an \textit{intrinsic} part
$A^{\rm sat}_{\rm intr}$ arising from the excitations created by the
sudden appearance of the core hole, and a term $A^{\rm sat}_{\rm
inter}$ from the \textit{interference} between them
\begin{equation} 
  A^{\rm sat} (\omega ,\omega')=A^{\rm sat}_{\rm ext} (\omega
  ,\omega')+A^{\rm sat}_{\rm intr} (\omega ,\omega')-2A^{\rm sat}_{\rm
  inter} (\omega ,\omega').
\end{equation} 
%Since the photoelectron and core screen each other, 
The effect of the
interference tends to reduce the satellite part of the spectral
function, and the spectral weight is shifted back to the
quasiparticle peak. This variation accounts for the $a(\omega )$
factor appearing in
the weight of the quasiparticle peak.  The detailed derivation of the
components of the spectral function arising from a PP dielectric
function have been presented elsewhere.\cite{campbell02} Here it is
sufficient to present results characterized by the extension to a
many-pole loss-function as in Eq.\ (\ref{epsqw}). Thus the intrinsic
and interference contributions are given by
\begin{eqnarray} 
  A^{\rm sat}_{\rm intr} (\omega ,\omega') & = & \frac{1}{\pi} \sum_i
  g_i \omega_i^2 \int_0^\infty \frac{dq}{\omega_i(q)^3}
  \delta(\omega'-\omega_i(q)), \\ A^{\rm sat}_{\rm inter} (\omega
  ,\omega') & = & \frac{1}{2\pi k} \sum_i g_i \omega_i^2 \int_0^\infty
  \frac{dq}{q\,\omega_i(q)^{2}} \delta(\omega'-\omega_i(q))\nonumber
  \\ & ~ & ~~~\times ~\ln \left[
  \frac{\omega_i(q)-q^2/2+k\,q}{\omega_i(q)-q^2/2-k\,q}\right], \\
  a(\omega) & = & \frac{1}{2\pi k_{0}} \sum_i g_i \omega_i^2
  \int_0^\infty \frac{dq}{q\, \omega_i(q)^{2}} \nonumber \\ & ~ &
  ~~~\times ~ \ln \left[
  \frac{\omega_i(q)+q^2/2+k_{0}\,q}{\omega_i(q)+q^2/2-k_{0}\,q}\right],
\end{eqnarray} 
where $k=[2(\omega -\omega')]^{1/2}$ is the photoelectron wavenumber
and $k_0=[2(\omega)]^{1/2}$ is the on-shell photoelectron wavenumber.
The extrinsic contribution to the spectral function can be found from
the photoelectron self-energy $\Sigma(k,\omega+\omega')$ and
renormalization constant $Z_{k}$.
\begin{eqnarray} 
  A^{\rm sat}_{\rm ext} (\omega,\omega') &\approx& -\frac{1}{\pi
  |Z_{k_{0}}|} \left[{ \frac{\Gamma _{k}+{\rm Im}\,\Sigma
  (k,\omega+\omega')} {[\omega' +\Delta_{k}]^{2}+[\Gamma _{k}]^{2}} }
  \right. \nonumber\\
  &&\ \ \ \ \ \ \ \ \ - \left.{ {\frac{{\rm Im}\,
  Z_{k_{0}}}{\omega' }} e^{-(\omega' /2\omega _{p})^{2}} }\right],
    \label{Aextr2} 
\end{eqnarray} 
where
\begin{eqnarray}
  \Delta_{k}&=& {\rm Re}\,[\Sigma (k_{0},\omega) -\,\Sigma
  (k,\omega+\omega')] \\
%\end{equation} 
%and 
%\begin{equation}
  \Gamma _{k}&=& -{\rm Im}\,[\Sigma (k_{0},\omega) - \Sigma
	 (k,\omega+\omega')].
\end{eqnarray}
% end Luke's part
%\section{Calculation}
%\label{CalcSect}
%Thus the method is quite efficient.

\section{XAS Calculations}
\label{ResultSect}

As illustrations of our approach, we now compare our results
for the XANES spectra of Cu and diamond as calculated with
the many-pole model against those calculated with the single-pole model,
and with experiment. Calculations of the self-energy
and the spectral function were converged with respect to the number
and distribution of poles used to represent the dielectric
function. We find that typically only $10 - 20$
poles are needed to represent a relatively broad loss function such as
that for Cu. The full multiple-scattering FEFF8 calculation for Cu was
converged with respect to the cluster size as well as the angular
momentum cutoff $l_{max}$. There are only two free parameters in our
calculations; a small imaginary shift in the potential was used
to account for experimental broadening; and a real energy shift
was introduced to correct for the inaccuracy in Fermi energies
calculated by the FEFF8 code.\cite{rehr06-2}

\begin{figure}[t] 
  \includegraphics[width=\columnwidth]{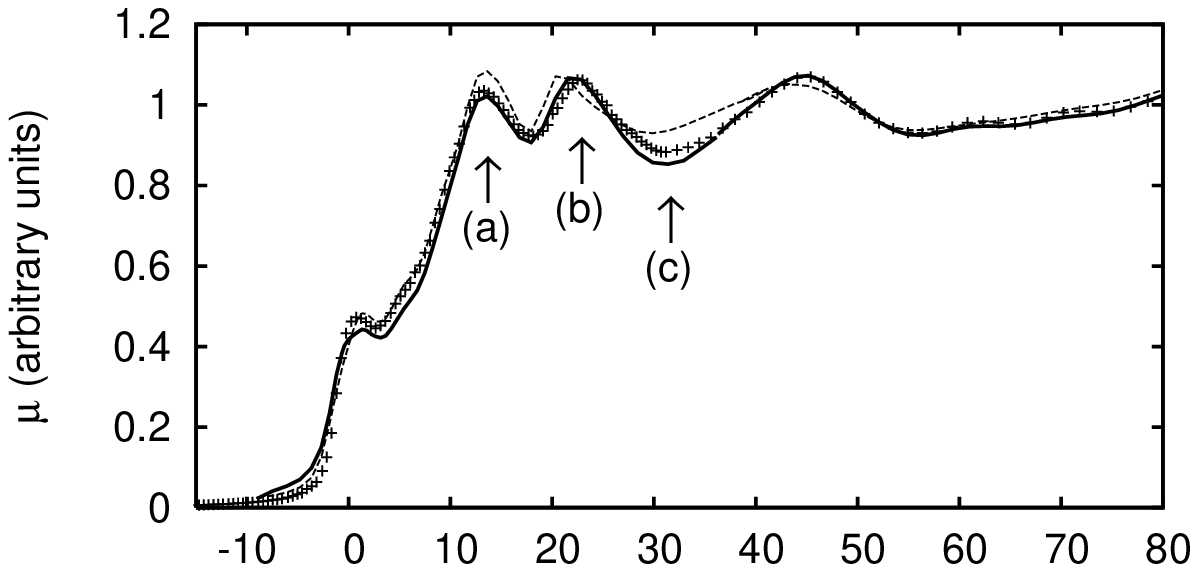}
  \includegraphics[width=\columnwidth]{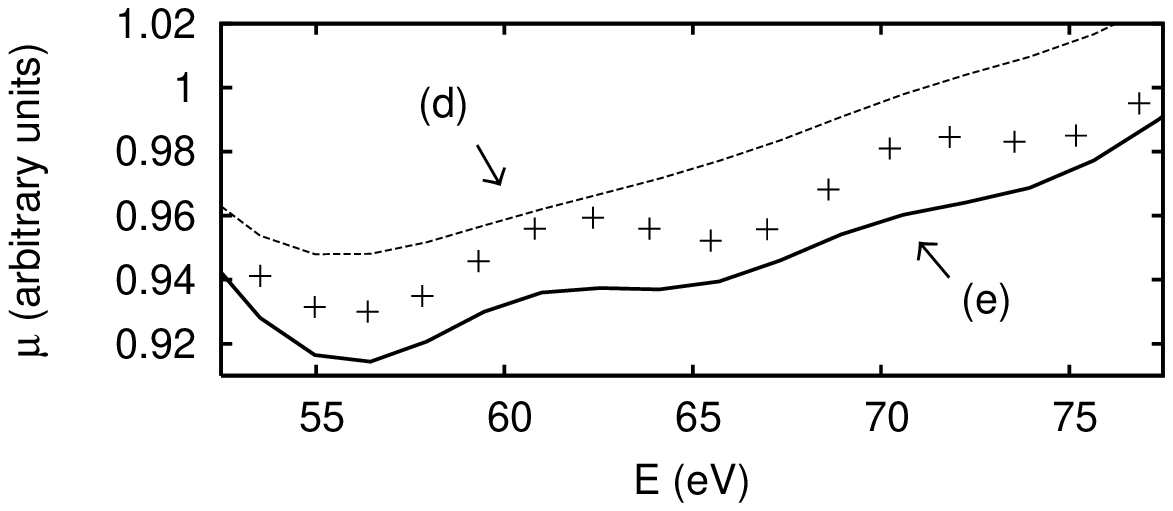}
  \caption{a) Top: Cu K edge XANES calculated from the many-pole self-energy
    and spectral function of this work (solid), and for comparison the
    conventional single-pole model (dashes), and experiment\cite{newville_pc}
(+); b) Bottom: Cu K edge EXAFS shown from $k=3.75-4.55 \AA^{-1}$.}
  \label{CuXanes}
\end{figure}

\begin{figure}[h]
  \includegraphics[width=\columnwidth]{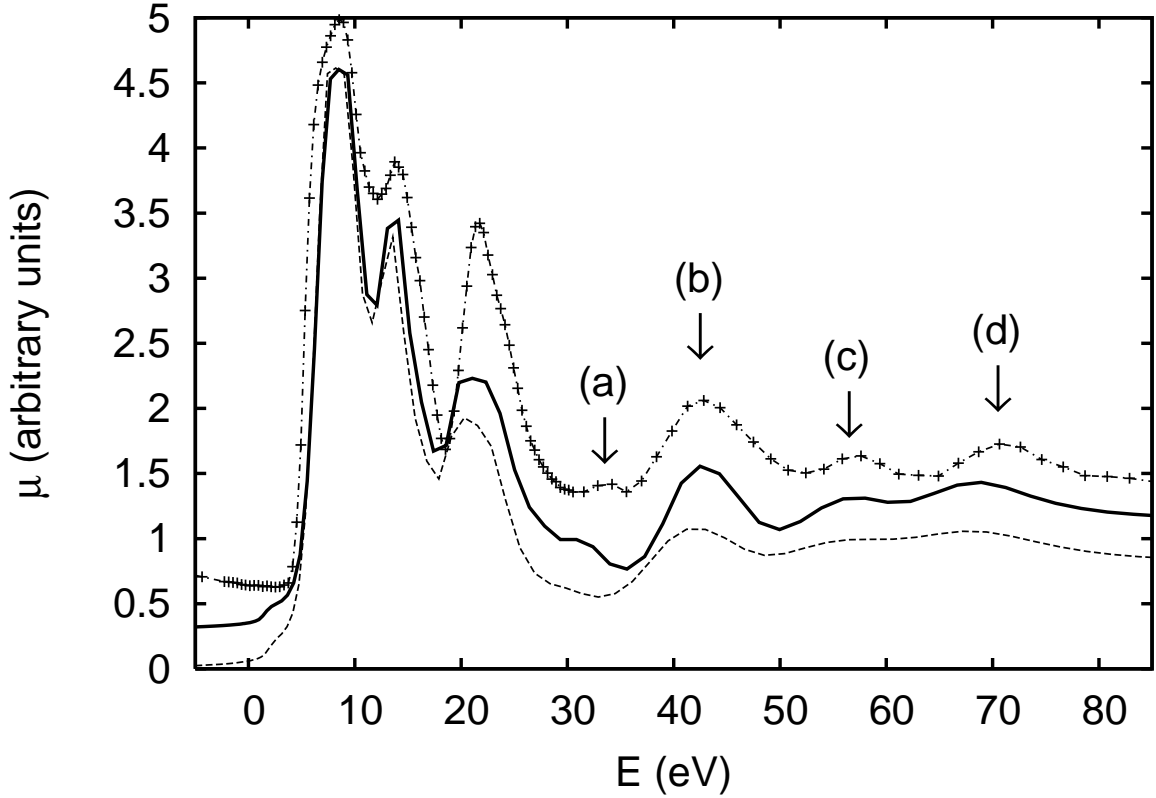}
  \caption{Diamond K edge XANES calculated with the many-pole
self-energy and spectral function of this work (solid), and for
comparison the conventional single-pole model (dashes), and experiment
(+).\cite{fister_07}}
  \label{DiaXanes}
\end{figure}

Comparison with experiment in Fig.\ \ref{CuXanes} and
\ref{DiaXanes} shows a clear improvement both in the phases and
amplitudes of the XAFS signal and the near-edge structure.  These
improvements can be linked respectively, to the real and imaginary
parts of the self-energy. The real part induces phase shifts in the
signal while the imaginary part is directly related to the inelastic
mean free path and hence the amplitudes.
Fig.\ \ref{CuXanes} shows our Cu K edge XANES calculations with both
single and many-pole self-energies compared to experiment. A large
(500 atom) cluster was used to calculate the spectra up to $\approx 35$ eV
above the Fermi level, above which a smaller (177 atom) cluster was
used with higher angular momentum components. Thus we used $l_{max} =
3$ for low energies and $l_{max} = 4$ above $\approx 35$ eV.  This was done in
order to ensure that errors due to finite cluster size and angular
momentum cutoff were small compared to effects of the self-energy on
the XANES spectrum. The result shows improvement in the amplitudes and
phases of the peaks, especially in the region from
$\approx 10-50$ eV (top). The amplitude of the ``whiteline'' peak
(a) is substantially reduced by the corrected self-energy, while the
second peak (b) acquires a phase shift. The dip seen at
$\approx 32$ eV (c) also attains a significant phase shift and an increase in
amplitude. The considerable improvements seen in this low energy XANES
region can be attributed to the fact that the plasmon pole self-energy has
singular behavior near the plasma frequency. This behavior is absent in
the many-pole self-energy which is naturally broadened by the width of
the loss function. Furthermore, there is improvement even in the EXAFS region
$~45-80$ eV (bottom). Here the single plasmon pole model gives a smooth,
almost featureless curve, whereas the many-pole model as well as the
experiment show noticeable features at $62$ eV (d) and $71$ eV (e).
%% Fig.\ \ref{AgXanes} is similar to Fig.\ \ref{CuXanes}, but for Ag K
%% XANES. In this case we found that 43 atoms with $l_{max}=4$ was
%% sufficient to converge the FEFF8 multiple-scattering
%% calculation. Again, amplitudes and phases are seen to be improved within
%% the first $40$eV. Beyond that, there is some discepancy in the phases,
%% which could be due to the finite angular momentum basis used in the
%% calculation.
%% \begin{figure}[t]
%%   \includegraphics[width=\columnwidth ]{Ag}
%%   \caption{Ag K edge XANES calculated from the many-pole self-energy
%% and spectral function of this work (solid), and for comparison the
%% conventional single-pole model (dashes), and experiment (+).\cite{bridges01}}
%%   \label{AgXanes}
%% \end{figure}
Fig.\ \ref{DiaXanes} presents similar calculations for the
diamond K edge XANES compared to data from non-resonant
inelastic x-ray scattering\cite{fister_07}. For diamond we could not fully converge the multiple
scattering calculations with respect to cluster size at all energies
because of memory requirements of the code. Thus we present our
results for a 500 atom cluster with $l_{max}=2$. Here the results are
more difficult to interpret because of errors due to finite cluster
size and our approximate treatment of core hole effects. However,
qualitative improvement is seen in the amplitudes of the
EXAFS from $\approx 25$ eV on. Specifically, the feature seen in the
experiment at approximately 32 eV (a) is absent in the single plasmon
pole calculation, but appears in the new calculation. Also, the three
subsequent peaks (b, c, and d) are enhanced as a result of the new
many-pole calculation, giving better qualitative agreement with
experiment. To reiterate, the single plasmon pole self-energy has a sharp
turn-on of the imaginary part which saturates too early, giving
excessive broadening in the range beyond the plasmon energy.
Similar self-energy effects have been seen in the F
K-edge spectrum of LiF, where a more computationally demanding full-GW
calculation was performed.\cite{soininen05}

\begin{figure}
  \includegraphics[height=\columnwidth,angle=-90]{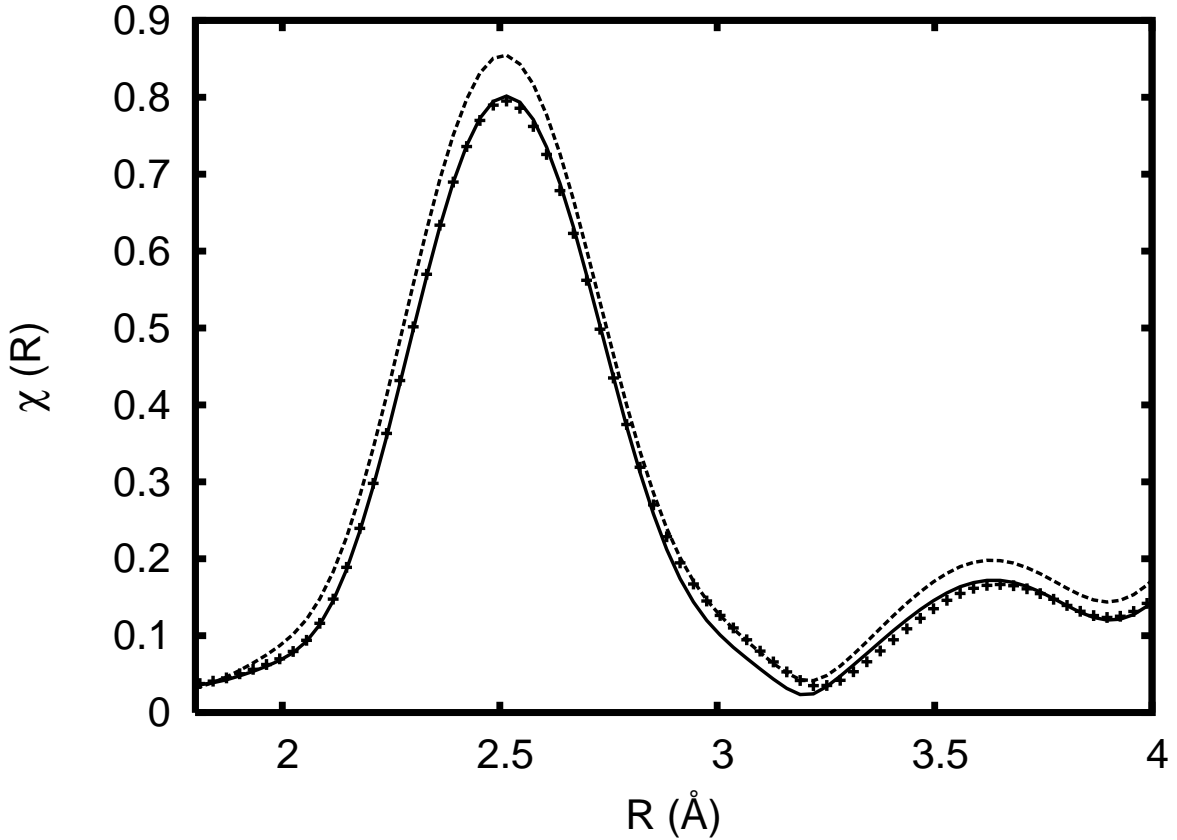}
  \caption{Comparison of Cu K edge EXAFS Fourier transform
    $\chi(R)$. Theoretical calculations using the many-pole self-energy
    (solid), the plasmon-pole model (dashed), and
    Experiment\cite{newville_pc} (+). The value of $S_{0}^{2}$ was found to be $\approx 0.93$ over most of the EXAFS range.} 
  \label{Cu_EXAFS}
\end{figure}

In addition to our XANES calculations, we have performed a comparison
of experiment and theory of the Cu K edge EXAFS using the analysis software
ATHENA.\cite{athena_artemis} To reduce Debye Waller effects, which are
highly correlated with the effects of the self-energy and many-body
amplitude reduction factor, we used data taken at a low temperature of
$10$ K. In order to give a fair comparison of the
two theories (PP self-energy and MPSE) we have fixed all
parameters to empirical or theoretical values.
First the theory and experiment were aligned by matching features in
the range $0-300$ eV. Then background subtraction and normalization was
performed using the same spline fitting range as well as normalization
range for experiment and theory. The EXAFS $\chi(k)$ was Fourier
transformed using a $k$-range of $2.632~\rm{\AA^{-1}}$ to $15.5~ \rm{\AA^{-1}}$
with a weighting of $k^1$. Debye waller factors were set using the
correlated Debye model with $\Theta_{D} = 315\rm{K}$. In addition, the
theory was broadened by $0.45$ eV half width half max to account for
experimental broadening. The estimate of the experimental resolution
was obtained by comparing to the width of the edge step. As can be
seen in Fig. \ref{Cu_EXAFS} the amplitude of the first shell peak is
reduced by our new treatment of inelastic losses, thus
improving the agreement with experiment and demonstrating the adequacy
of our calculation of $S_0^2$. Our value for $S_0^2$ ($\approx 0.93$)
also agrees with a crude approximation (previously implemented in the
FEFF8 code) which calculates the many-body overlap of the atomic system
and gives $S_0^2 = 0.95$.

%For example, a comparison of fits to the Ag K-edge spectrum give
%results for the R-Factor which are slightly better when using the%many-pole self-energy than the those obtained using the plasmon-pole
%self-energy. In addition, the physical parameters are consistent
%between the two models.

\section{Conclusions}\label{ConcSect}

An efficient many-pole model for calculations of inelastic losses
has been developed and successfully implemented in an extension of the
multiple scattering code FEFF8.
Our many-pole model is based on an \textit{ab initio} calculation
of the zero momentum transfer loss function by means of the
RSGF approach implemented in an extension of the FEFF8 code.
Extrapolation to finite momentum transfer is performed by representing
$\epsilon({\bm q},\omega)^{-1}$ as a sum of poles.
The approach yields both the  
quasiparticle self-energy to account for extrinsic losses
and the many-body amplitude factor $S_0^2$ to account for
intrinsic losses and interference terms.
The validity of the self-energy model
was checked by comparison with more detailed, first principles
calculations.\cite{soininen03}
We find that $S_0^2$
is nearly energy independent over a broad range.
%The approach yields improved calculations of x-ray absorption spectra
%and related spectroscopies such as electron energy loss spectra (EELS),
%most notably in the near edge spectra.
%In addition, the many-pole self-energy
%model for extrinsic losses has been extended to incorporate
%%many-body effects (including intrinsic and interference effects in
%calculations of XAS via a
%convolution with the quasiparticle spectral function.
Calculations with the
many-pole model are shown to improve agreement with experimental
results for the near edge XAS of several materials. In addition we
find that our model is consistent with the plasmon-pole model when
applied to the extended (EXAFS) region, which is an important step
toward quantitative full spectrum calculations.
A drawback of the present model is that it does not fully
%The model is lacking in the fact that
account for the contribution due to the particle-hole continuum at low
energies. % though this could be added in the future.
Thus the current approach may be expected to give better results for materials
with broad loss functions, since in these cases the self-energy will
be dominated by plasmon-like excitations even at low energies. Other
improvements would be to represent the energy loss
function as a sum of broadened poles with momentum transfer dependent
broadening, and to better account for the particle-hole  continuum;
efforts along these lines are in progress.  Finally we note that
the development of {\it ab initio} calculations of inelastic losses
here, together with the recently developed \textit{ab initio}
Debye Waller factor calculations\cite{vila07}
yields improved first principles calculations of XAS from
structural coordinates alone, without phenomenological models or
the need to fit theoretical model self-energy, mean free path,
or many-body amplitude parameters.

%Notes: 1) The self-energy in EXAFS is about the same - i.e., PP
%fine in EXAFS.  2) combined with ab initio DW factors, now all
%quantities in EXAFS eq are calculated from first principles.
%3) Extend to finite q
%4) Calculate G for system e.g., with MS expansion.
%5) Add local fields
\begin{acknowledgments}
This work was supported by DOE Grant DE-FG03-97ER45623 (JJR,MPP), NIH NCRR
BTP Grant RR-01209 (JJK), NIST Grant 70 NAMB 2H003 (APS), and
Academy of Finland Contract No 201291/205967/110571 (JAS) and was
facilitated by the DOE Computational Materials Science Network.
\end{acknowledgments}

\appendix
\section{Plasmon-pole self-energy }
\label{PPApp}
Here we give a more complete description of the plasmon-pole (PP)
model of Hedin and Lundqvist.\cite{BIL67_I,BIL67_II,hedin69}
We begin with the
GW approximation for the self-energy\cite{hedin69} 
%of Eq.~(\ref{SPEpsEq}),
\begin{equation}
  \Sigma(\bm{r},\bm{r}',E) = i \int \frac{d\omega}{2 \pi}\,
  G(\bm{r},\bm{r}',E-\omega)W(\bm{r},\bm{r}',\omega).
\end{equation}
Here $G$ is the single-particle Greens function, which has a spectral
representation
\begin{equation}
  G(\bm{r},\bm{r}',E) =
  \sum_i {\frac{\phi_i(\bm{r})\phi^{*}_i(\bm{r}')}
         {E-E_i + i\delta\, \sgn(E_{i}-E_{F})} },
\end{equation}
and $W$ is the dynamically screened coulomb potential,
\begin{eqnarray}
  W(\bm{r},\bm{r}',\omega) & = & \int d^3r''\,
  \epsilon(\bm{r},\bm{r}'',\omega)^{-1}V(\bm{r}'',\bm{r}'), \\
  V(\bm{r},\bm{r}') & = &\frac{1}{|\bm{r}-\bm{r}'|}.
\end{eqnarray}
Here $V$ is the bare Coulomb potential and $\epsilon^{-1}$ is the
inverse dielectric matrix. 
Using the Green's function for a homogeneous electron gas, the self-energy
in the momentum representation is given by
Eq.~(\ref{SPSE_Eq}).
%\begin{equation}
%  \label{SPSE_Eq}
%  \begin{split}
%    \Sigma(\bm{k},E) = &\, i \int \frac{d^3q}{(2
%    \pi)^{3}}\,\frac{d\omega}{2 \pi}
%    \frac{V(q)}{\epsilon(\bm{q},\omega)}\\ \times
%    &\frac{1}{E-\omega-E_{\bm{k}-\bm{q}}
%    +i(|\bm{k}-\bm{q}|-k_{F}) \delta},
%  \end{split}
%\end{equation}
%where $k_F$ is the Fermi momentum.
In frequency space the imaginary
part of the inverse dielectric function (i.e., the loss function of
the electron gas) is modeled as a single pole at $\omega(q) =
[\omega_p^2 + a q^2 + b q^4]^{1/2}$, where the coefficients of the
dispersion $a=k_{F}^{2}/3$ and $b=1/4$ are chosen to give the
Thomas-Fermi potential at low frequency, as well as the correct high
momentum transfer limit.\cite{BIL67_II,hedin69} This gives an
inverse dielectric function whose imaginary part is given by
\begin{equation}
  \label{SPEpsEq2}
L({\bm q},\omega) =
 -{\rm Im} \left[\epsilon(\bm{q},\omega)^{-1}\right] = \pi \omega_p^2\,
  \delta[\omega^2-\omega(q)^{2}].
\end{equation}
The real part of the loss function can be obtained
via a Kramers-Kronig transform
\begin{eqnarray}
%  \begin{split}
    {\rm Re} \left[\epsilon(\bm{q},\omega)^{-1}\right] &=& 1 -
    \frac{1}{\pi} \int_0^{\infty} d\omega'\, \frac{2
    \omega'}{\omega^2-\omega'^2}\, {\rm
    Im} \left[\epsilon(\bm{q},\omega')^{-1}\right] \nonumber \\
&=& 1 + \frac{\omega_p^2}{\omega^2-\omega(q)^2}.
%  \end{split}
\end{eqnarray}
Inserting these results into Eq.\ (\ref{SPSE_Eq}) then yields two
terms: the first term can be integrated analytically and gives a
static Hartree-Fock exchange potential $\Sigma_{HF}$ 
\begin{equation}
  \Sigma_{HF}(k) = -\frac{k_F}{\pi}\left[1 + \frac{k_F^2 - k^2}{2 k
      k_F}\ln\left|\frac{k_F+k}{k_F-k}\right|\right].
\end{equation}
The second term, denoted by $\Sigma_d({\bm k},E;\omega_p)$, is the
dynamically screened exchange-correlation contribution, which can be
interpreted as the one loop diagram containing the electron propagator
$G$ and a boson (plasmon) propagator
\begin{equation}
  D(\bm{q},\omega) = \frac{2 \omega(q)}{\omega^2 - \omega(q)^2 +
    i\delta}.
\end{equation}
Thus the dynamic term $\Sigma_d$ arises from the creation of virtual
bosons which interact with the photoelectron via an effective coupling
$|g(q)|^2 = \omega_{p}^{2} V(q)/2 \omega(q)$.
%We define this second term as $\Sigma_d(\bm{k},E)$.
The integral over $\omega$ and solid angle in Eq.\ (\ref{SPSE_Eq}) can
be done analytically, so that $\Sigma_d(\bm{k},E)$ is given by a
one-dimensional integral over $|\bm{q}|$.  The resulting expression for
the self-energy is quite lengthy and is not reproduced here,
but can be found in Eq.\ (13) of Ref. \onlinecite{BIL67_II}.
% p. 208.
%\input{HLQEq.tex}
\section{Equivalence of Self-Energy Formulae}
\label{HLQEqApp}
%\begin{figure}
%  \label{ContourFig}
%  \includegraphics[width=3 in]{contours}
%  \caption{Contours of integration (shown in gray) for the situation
%    where the pole of the Greens function is located in quadrant 1, 2,
%    3 or 4.} 
%\end{figure}
In this appendix we demonstrate that self-energy expressions of Hedin and
Lundqvist and of Quinn and Farrell are essentially equivalent,
except for slight differences the approximations used.
We start with the self-energy of an electron gas
within the $GW$ approximation as
given by Hedin and Lunqvist\cite{BIL67_I} in Eq.~(\ref{SPSE_Eq}).
%\begin{equation}
%  \label{SPSE_Eq}
%  \begin{split}
%    \Sigma(\bm{k},E) = &\, i \int \frac{d^3q}{(2
%    \pi)^{3}}\,\frac{d\omega}{2 \pi}
%    \frac{V(q)}{\epsilon(\bm{q},\omega)}\\ \times
%    &\frac{1}{E-\omega-E_{\bm{k}-\bm{q}}
%    +i(|\bm{k}-\bm{q}|-k_{F}) \delta}.
%  \end{split}
%\end{equation}
This expression can be split into two terms.
The Hartree-Fock exchange potential
$\Sigma_{HF}$, and the dynamically screened exchange-correlation
potential $\Sigma_{d}$ which includes the dynamic response
proportional to $[\epsilon(\bm{q},\omega)^{-1} - 1]$.
\begin{equation}  
  \Sigma(\bm{k},E) = \Sigma_{HF}(\bm{k}) + \Sigma_{d}(\bm{k},E)
\end{equation}
and
\begin{equation}
  \label{SigmadEq}
  \begin{split}
    \Sigma_{d}(\bm{k},E) = &\, i \int
    \frac{d^3q}{(2\pi)^{3}}\,\frac{d\omega}{2 \pi} 
    V(q)\left[\epsilon(\bm{q},\omega)^{-1} -1\right]\\ \times
    &\frac{1}{E-\omega-E_{\bm{k}-\bm{q}}
      +i(|\bm{k}-\bm{q}|-k_{F}) \delta}.
  \end{split}
\end{equation}
If we rewrite $\epsilon(\bm{q},\omega)^{-1}$ in its spectral
representation
\begin{equation}
  \epsilon(\bm{q},\omega)^{-1} = 1 - 1/\pi \int d\omega'
  \frac{2\omega'}{\omega^{2}-(\omega'-i\delta)^2}
  {\rm Im} \left[\epsilon(\bm{q},\omega)^{-1}\right],
\end{equation}
Eq.~(\ref{SigmadEq}) becomes
\begin{equation}
  \begin{split}
    \Sigma_{d}(&\bm{k},E) = -\frac{i}{\pi} \int_{0}^{\infty} \int
    \frac{d^3q}{(2\pi)^{3}}\, 2 \omega' V(q) {\rm Im} \left[\epsilon(\bm{q},\omega')^{-1} -1\right]
    \\ \times &
    \int\frac{d\omega}{2 \pi} \frac{1}{\omega^{2}-(\omega'-i\delta)^{2}}
    \frac{1}{E-\omega-E_{\bm{k}-\bm{q}}
      +i(|\bm{k}-\bm{q}|-k_{F}) \delta}.
  \end{split}
\end{equation}

The integral over $\omega$ can be performed by deforming the contour to
the imaginary axis and including residues of the Greens function when
necessary.
%as shown in Fig. \ref{ContourFig}.
The integral along the
imaginary axis is purely real, thus the imaginary part of the
self-energy is given by the imaginary part of the contribution from
the residues of the poles in the Greens function. The result can then
be split into two terms. One arises from the particle contribution
and occurs for energies greater than the Fermi energy
\begin{eqnarray}
%  \begin{split}
    {\rm Im}\,\left[\Sigma({\bm k},E)\right] &= & \int
    \frac{d^{3}q}{(2\pi)^{3}}
    \Theta\left({\rm Im} \left[ \Delta E_{{\bm k}-{\bm q}} \right] \right)
    \Theta\left({\rm Re} \left[ \Delta E_{{\bm k}-{\bm q}} \right]\right)
\nonumber \\
    &\times & \frac{1}{q^{2}} 
    {\rm Im} \left[\epsilon({\bm q}, \Delta E_{{\bm k}-{\bm q}})^{-1}\right],
%  \end{split}
\end{eqnarray}
where
\begin{equation}
    \Delta E_{{\bm k}-{\bm q}} = E-E_{{\bm k}-{\bm q}} + 
    i\delta(|{\bm k}-{\bm q}|-k_{F}.
\end{equation}
The other is associated with the hole contribution where the energy
is less than the Fermi energy
\begin{eqnarray}
%  \begin{split}
    {\rm Im}\, \left[\Sigma({\bm k},E)\right] & = & -\int
    \frac{d^{3}q}{(2\pi)^{3}}
    \Theta\left(-{\rm Im} \left[ \Delta E_{{\bm k}-{\bm q}} \right] \right)
    \Theta\left(-{\rm Re} \left[ \Delta E_{{\bm k}-{\bm q}} \right]\right)
    \nonumber \\
   & \times & \frac{1}{q^{2}} 
    {\rm Im} \left[\epsilon({\bm q}, \Delta E_{{\bm k}-{\bm q}})^{-1}\right].
%  \end{split}
\end{eqnarray}
Quinn and Ferrell make the further approximation that
$E = k^{2}/2$, which yields the formula derived in
Ref. \onlinecite{quinn_58} and is
used as a starting point by Penn.\cite{penn87,quinn_62} Thus, Penn's
formulation is equivalent to that of
Hedin and Lundqvist\cite{BIL67_I,BIL67_II} with zeroth order
approximations for the quasiparticle energy and
renormalization constant.
%In the LDA, some
%of these variables are related to the electron density $n=N/V$.
%\begin{equation}
%  \omega_{p} = (4 \pi n)^(1/2) ,\qquad k_F = (3 \pi^{2} n)^(1/3).
%\end{equation}
%approximation some of these variables are related
%$\Sigma(E)$ can also be expressed as a function of these variables
%$E, k$, and $\rho$ since
%and
%\begin{equation}
%\end{equation}
%by writing the position dependent self-energy as
%\begin{equation}
%  \Sigma(\bm{r},\bm{r}') =
%  \Sigma(k(E),E,\rho(\bm{r}))\delta(\bm{r}-\bm{r}')
%\end{equation}
%where k(E) is found by solving for the quasiparticle momentum 
%\begin{equation}
%  k(E) = \sqrt{2 (k^{2}/2 + \Sigma(k(E),E)) }
%\end{equation}
%self consistently.
% especially for systems where the valence or conduction
%electrons are described adequately by an electron-gas
%model. Consequently the model is well suited for nearly free electron
%metals such as aluminum, with a sharp plasmon peak in the inverse
%dielectric function. 

\end{document}